# The Impact of Attending a Remedial Support Program on Syrian Children's Reading Skills: Using BART for Causal Inference

Mayarí I. Montes-de-Oca, Jennifer L. Hill,
Lawrence Aber, Carly Tubbs Dolan,
Kalina Gjicali

*New York University*

Data collection and processing were funded by the Spencer Foundation, Dubai Cares, and the E-cubed Research Fund. Funding for scientific analysis and write-up was provided by the New York University Abu Dhabi and the E-cubed Research Fund.

**Contact Information**

Mayarí Montes de Oca, mmd486@nyu.edu

Jennifer L. Hill, Department of Applied Statistics, Social Science, and Humanities, NYU, jennifer.hill@nyu.edu

Lawrence Aber, Department of Applied Psychology, NYU, la39@nyu.edu

Carly Tubbs Dolan, NYU Global TIES for Children, carly.tubbs@nyu.edu

Kalina Gjicali, NYU Global TIES for Children, kg1317@nyu.edu

**Abstract**

This article estimates, for a sample of 1,777 Syrian refugee children, the impact on basic reading assessments of attending a remedial support program in Lebanon that was infused with social and emotional learning practices. We use flexible methods that capitalize on advantages of both machine learning and Bayesian inferential frameworks to leverage the information available in understudied contexts and help account for the problem of self-selection. Average treatment effects were estimated both using multiply imputed data and data from outcome-respondents only. We do not find conclusive evidence for an effect on one of the reading measures studied (ASER). However, we provide evidence for positive effects for three, more robust, measures of basic reading outcomes from the Arabic EGRA assessment. We discuss potential reasons for the differences in effects that are relevant for educational research and practice. We also consider the implications for future research of choices related to measurement, data collection and processing, and missing data.

***Keywords*****:** causal inference, refugees, social-emotional learning, program evaluation, attendance, nonparametric, Bayesian, non-formal education, selection bias.

## Introduction

6.6 million Syrians have fled their home country since 2011, mostly to neighboring countries of Turkey, Lebanon, Jordan, Iraq, and Egypt, as a result of a protracted violent conflict (UNHCR, 2021). Currently, there are at least 855,172 Syrian refugees in Lebanon, 90% of whom live on less than $2.90 USD per day and 512,500 of whom are estimated to be children and youth (UNHCR, 2020).

Refugee children have dramatically lower enrollment rates in school than host country children on average (UNHCR, 2019; Tubbs, Kim, Brown, Gjicali, & Aber, 2021). Moreover, among those enrolled, chronic absenteeism is prevalent (Sieverding, Krafft, Berri, Keo, & Sharpless, 2018; Tumen, Vlassopoulos, & Wahba, 2021; Brown, Kim, Tubbs, & Aber, 2021). This combination of circumstances poses the risk of profoundly hindering the socioemotional and academic development of a full generation of children and youth.

As a response to this crisis, several efforts by local and international agencies have been put in place to mitigate the consequences of displacement and armed conflict on the development of these children (UNHCR, 2012; The World Bank, 2018; Tubbs et al., 2021). The Lebanese Ministry of Education and Higher Education, for example, has committed to increasing access to Lebanese public schools for Syrian refugee students (MEHE, 2016; Kim, Brown, Tubbs, Sheridan, & Aber, 2020). This has placed substantial pressure on the Lebanese education system -- a system unprepared to address the specific socioemotional and academic needs of Syrian refugees (Kim et al., 2020).

Syrian refugee students face a different reality from that of national Lebanese students. Differences include language adaptation challenges, different grade placement due to previous school interruptions (Shuayb, Makkouk, & Tuttunji, 2014; Kim et al., 2020), food insecurity, as well as unstable and overcrowded living conditions (UNHCR, 2020). Moreover, refugee students often face cultural and political tensions with their host communities, students, and teachers (UNHCR, 2013; Yoshikawa, Wuermli & Aber, 2019).

**Remedial Programing And Social And Emotional Learning**

While more efforts from the international community have been put in place to integrate refugee children into local education systems, evidence regarding which strategies improve refugee children's learning outcomes is still scarce. Prior studies suggest that access to formal schooling does not always translate into improved learning outcomes (Pritchett, 2015; Torrente et al., 2015; Aber, Tubbs, et al., 2017).

After-school remedial programs have played a central role in the efforts launched by INGO's and local governments to mitigate the learning crisis that ongoing armed conflicts have left behind (Jalbout, 2015). Particularly, social and emotional learning (SEL) programming, both in school and after school, has been increasingly incorporated into the educational response. The goal is to provide children in conflict-affected contexts with psychosocial support to mitigate the effects that migration, conflict, and trauma can have on their academic development (Aber, Torrente, et al., 2017; UNESCO, 2019; Yoshikawa et al., 2019; Tubbs et al., 2021; Brown et al., 2021). These programs often involve providing teachers with training and support to promote safe and supportive learning environments that can help improve the students' self-perception, social and self-awareness, interpersonal skills, self-management and regulation of their emotions, among other skills (Denham & Brown, 2010; Tubbs et al., 2021). Numerous studies evaluating in-school and after-school SEL programs in the United States and in other Western countries have found positive impacts on student's academic performance (Jones, Brown, Hoglund, & Aber, 2010; Durlak, Weissberg, Dymnicki, Taylor, & Schellinger, 2011; Schonfeld et al., 2015) and emerging evidence suggests that offering ***access*** to SEL education programming could improve student's learning and development in crisis contexts (Aber, Torrente, et al., 2017; Aber, Tubbs, et al., 2017; Tubbs et al., 2021; Brown et al., 2021; Aber, Tubbs, Kim, & Brown, 2021). Nonetheless, there is limited evidence on how to implement such programs effectively (Torrente et al., 2019) under conditions that compromise children's ability to regularly attend. The amount of exposure to an SEL intervention during a trial can vary dramatically due to numerous reasons, including heterogeneity of the days offered in each classroom, security threats, and most importantly due to significant differences in the children's circumstances and risk factors. The consequences of this non-random differentiated exposure remain understudied and, if ignored, the potential benefits that remedial support and SEL can have on the academic development of refugee children cannot be fully understood.

**Attendance**

Despite the efforts and emergent research around increasing *access* to schooling in conflict-affected contexts*, attendance* in the classroom is "voluntary" and often driven by financial hardships, migration patterns, and household instabilities. Prior research from developing countries and from other emergency contexts has documented the prevalence of absenteeism among at-risk children, and the challenges that it represents for international efforts that target their academic development (Banerjee et al., 2010; Arbour et al. 2016; Tubbs et al., 2021). Moreover, absenteeism from regular schooling in Syrian refugee children has been found at alarmingly high rates (Sieverding et al., 2018; Tumen et al., 2021), and attendance at *out-of-school* remedial programs is expected to be more difficult to attain and sustain. While it would be helpful to better understand the impacts of attendance/absenteeism on subsequent outcomes, differential selection into levels of attendance complicates this estimation.

Evidence from the implementation of remedial programming in another conflict-affected context, Niger, suggests that providing *access* alone cannot leverage all the benefits that low-cost targeted remedial programming can have on children's learning outcomes (Brown et al., 2021). Results from a study in Chile reflects the important role that attendance plays in moderating the effects of educational interventions for at-risk children. In fact, ignoring attendance as a moderator can lead to the misleading conclusion that such interventions have no impact (Arbour et al., 2016).

Furthermore, compelling correlational and regression-based studies have illustrated the importance of the relationship between attendance and learning outcomes, and of the role that absenteeism plays as a risk factor in children's academic development (Finn, 1993; Lamdin, 1996; Maclean et al., 2016; Gershenson et al., 2017; Tubbs et al., 2021).

Fewer studies, however, have focused on the study of the *impact* of attendance on academic outcomes under a causal framework. Evidence from studies in the United States, Australia, and Ghana has supported the reasonable hypothesis that attendance matters in children's learning and that it affects academic outcomes (Cannon, Jacknowitz, & Painter, 2006; Gottfried, 2010, Goodman, 2014; Opoku & Boahen, 2021). Some studies suggest that the benefits of attendance, however, may not be achieved by the most at-risk participants living under the poverty line (Cannon et al., 2006).

The context and challenges that Syrian refugee populations face are without a doubt atypical, heterogeneous, and constantly shifting, and they remain understudied. Furthermore, attendance in *out-of-school programs* has the potential to have different implications for learning outcomes than attendance in regular schooling. Creating a better understanding of attendance-outcome relationship of interventions that target the well-being of these populations, while attempting to address the challenges it presents to causal inference, can help us assess whether remedial education holds promise for refugee and conflict-affected students.

**Current Study**

As part of the efforts to address the specific needs of refugee children, the International Rescue Committee (IRC) and NYU Global TIES for Children have conducted and evaluated a series of interventions that provide remedial tutoring programs for refugee children who are enrolled in public schools in their host communities (IRC, 2021). The present study uses a longitudinal sample of Syrian refugee children in Lebanon who participated in *one* of these cluster-randomized control trials in order to study the impact of *attending* non-formal remedial education on children's learning.

Most impact evaluation studies perform intent-to-treat analyses that ignore to what degree participants actually received the treatment. To determine the amplitude of effectiveness of programming, this analysis considers students' *attendance* as a measure of dosage. By focusing on variability in program uptake we aim to better understand the impact that interventions can have on children's learning and development.

**Research Question**

The children studied in this paper were offered access to non-formal remedial instruction infused with SEL principles and followed over the course of 26 weeks during the academic year 2016-2017, to address the following research question:

**RQ.** What is the *impact* of **attendance** to non-formal SEL-infused remedial programming on refugee children's basic reading outcomes?

However, attendance in the classroom was voluntary and not randomized, and in order to identify causal effects under these circumstances special methods and stronger assumptions are needed, than if we could ensure a successful randomization to the program's exposure. We therefore use innovative methods for causal inference in the presence of self-selection.

**Main Contributions**

The main contributions of this study lie at the intersection of (1) an understudied crisis-context and population, (2) the understudied impact of attendance in *out-of-school* SEL programming, (3) the rigor to explicitly address and assess how *missing data* affected our study, and (4) the study of attendance through *flexible methods for causal inference* that (a) capitalize on the advantages of machine learning and statistical frameworks to better account for the problem of self-selection and that (b) can leverage a large amount of information in an understudied context.

***Understudied context and population.*** As a response to the lean evidence base on the conditions, challenges, and developmental processes of Syrian refugee children, the Education in Emergencies: Evidence for Action (3EA) partnership between NYU Global TIES for Children and the IRC collected an extensive amount of information on the children and their families, their migration history, the teacher's background characteristics and instructional practices, and on their communities. The data collection process was driven by a team of measurement and child development experts and the resulting dataset provides hundreds of potentially relevant confounders and covariates. One of the distinctive features of this study is that, to leverage such a rich data set and to limit generalizations from Western countries to Syrian refugee populations, we use methods that capitalize on the ability of machine learning algorithms to estimate relationships between attendance and outcomes while adjusting for numerous covariates (Hill, Weiss, & Zhai, 2011; Degenhardt, Seifert, & Szymczak, 2019).

***Relaxing parametric assumptions.*** Accounting for many covariates, however, presents a modeling challenge, particularly in a poorly understood context where we want to avoid making strong assumptions about the precise forms of the relationships between confounders and the outcome, as well as about the potential interactions with the treatment. In contrast to linear models, the methods used in the present analysis account for such complexities naturally, reducing the amount of parametric assumptions that need to be made (Hill, 2011; Hill, Linero, & Murray, 2020). This minimizes the risk of bias from excessive tweaking of functional forms and limits the introduction of cultural bias from prior Western-based research results.

***Common support for valid causal inference claims.*** Furthermore, the methods applied here make explicit use of the potential outcomes framework (Rubin, 1987, 2005) and offer principled ways to assess whether there is enough empirical support to estimate counterfactual states (and to, thus, make causal claims) at different levels of attendance and for each participant. We draw from the common support approaches developed by Hill and Su (2013) to assess the availability and quality of empirical counterfactuals in our sample. In addition, we use the Average Dosage-Response Function framework for causal inference (Galagate, 2016) to account for differences in the causal relationship between attendance and the outcome across participants.

***Instrumental variables.*** Many studies use instrumental variables (IV) approaches to address these questions. We see several downsides to this approach. First, IV requires that the exclusion restriction is satisfied. In this context, this means that the only pathway by which randomization to the treatment can affect student outcomes is through attendance. However, alternate pathways are possible. The child's and parents' general engagement with their academic development can be affected as a result of being granted access to the program, and factors such as the remedial teachers’ attendance and engagement could affect students' outcomes, regardless of the students' individual attendance. Moreover, using a binary instrument to identify a continuous or multi-valued treatment is an extension of the basic IV framework that requires stronger assumptions about the functional form of the model than the simple binary-instrument, binary-treatment context. Traditional IV models are designed to estimate various forms of local average treatment effects and are not generally capable of identifying dose-specific effects. Finally, instrumental variables methods typically ignore issues of *common support* that, in contrast, the BART-based methods are designed to diagnose. The fully parametric IV regression-based approaches, don’t provide ways to assess the availability of adequate counterfactuals at *different levels* of attendance. Rather, they are concerned with the impact of moving from zero to the observed, non-random, levels of attendance. This means that they have to make assumptions about the relationship between attendance and the outcome for participants with observed low-attendance, based on the outcome of participants with observed high-attendance, and vice versa. Which risks overconfident extrapolation of estimations to participants for whom we may lack empirical support for causal inference claims.

***Addressing missing data.*** Missing data, which is prevalent in the study of crisis-affected migrant populations, is often either ignored or not given sufficient attention. In this study we devote efforts to 1) examine non-response patterns and diagnostics, 2) address missing data in a principled way, 3) compare how impact estimates differ before and after adjustment for non-response, and 4) thoroughly discuss its implications in research findings.

**Healing Classrooms**

The intervention of focus in this study is a SEL-infused remedial academic support program implemented by the IRC, referred to as “Healing Classrooms”. The program focused on improving teachers’ ability to create a supportive environment during their delivery of basic literacy and numeracy instruction. It was delivered by the IRC at community-based sites recruited in the Akkar and Bekaa regions of Lebanon. Randomization to access occurred at the site level, before the academic year 2016-2017, and attendance varied greatly across students after enrollment.

The IRC Healing Classrooms programs focus on the role that teachers play in promoting the psycho-social recovery and well-being of students. They aim to foster an inclusive approach to education, with awareness of the impact that violence and displacement have on children and of the increased risks that they face. They do this by offering training, ongoing support, monitoring and referral systems (Child and Youth Protection and Development Unit, 2011). As part of these programs teachers participate in Teacher Learning Circles (TLCs), and receive mentoring and training to promote in children a sense of self-worth, self-control, belonging and, more generally, positive social skills (Child and Youth Protection and Development Unit, 2011).

In recent years the 3EA partnership conducted evaluations of the access to these programs in the Democratic Republic of the Congo and in Niger. Previous evidence from these evaluations suggests that being offered *access* to Healing Classrooms helped increase students' perceptions of the supportiveness of teachers and schools, as well as their academic outcomes (Torrente et al., 2015, 2019; Aber, Torrente, et al., 2017; Aber, Tubbs, et al., 2017; Brown et al., 2021).

**Conceptual Framework**

The overarching conceptual framework for this study is based on educational and developmental theory. We draw on "developmental-contextual models" which view development as taking place in a nested, interactive set of contexts, ranging from immediate (e.g., family, peer system, classroom) to more distal (e.g., cultural and political) (Bronfenbrenner & Morris, 1998; Cicchetti & Aber, 1998). The classroom serves as the context of focus in this study. Children's development depends on the classroom, which provides opportunities for interactions through relationships that support development. In particular, in our context, the "Healing Classrooms" (HC) infuses SEL principles into instruction (Aber, Torrente, et al., 2017). The conceptual framework proposes that for children to effectively acquire SEL and academic skills, they need to participate in safe and high-quality learning environments. Healing Classrooms supports this through positive teacher-child relationships, which have been shown to be associated with social competence and positive school adjustment (Baker, 2006; Hamre & Pianta, 2001; Pianta & Stuhlman, 2004; Rimm-Kaufman & Hamre, 2010).

Figure 1 illustrates the main components of this framework. We consider that the uptake (attendance, engagement) and high-quality delivery of the program are crucial for the benefits of SEL-infused programming to affect academic outcomes. We also assume that pre-existing characteristics of the child at the individual, family, and community level, can affect the academic outcomes and uptake levels, and play a role in moderating the impacts of the program uptake. The key elements for the present analysis are: (1) child attendance in remedial programming (a proxy for program uptake); and (2) child reading outcomes.

**Figure 1** Conceptual framework

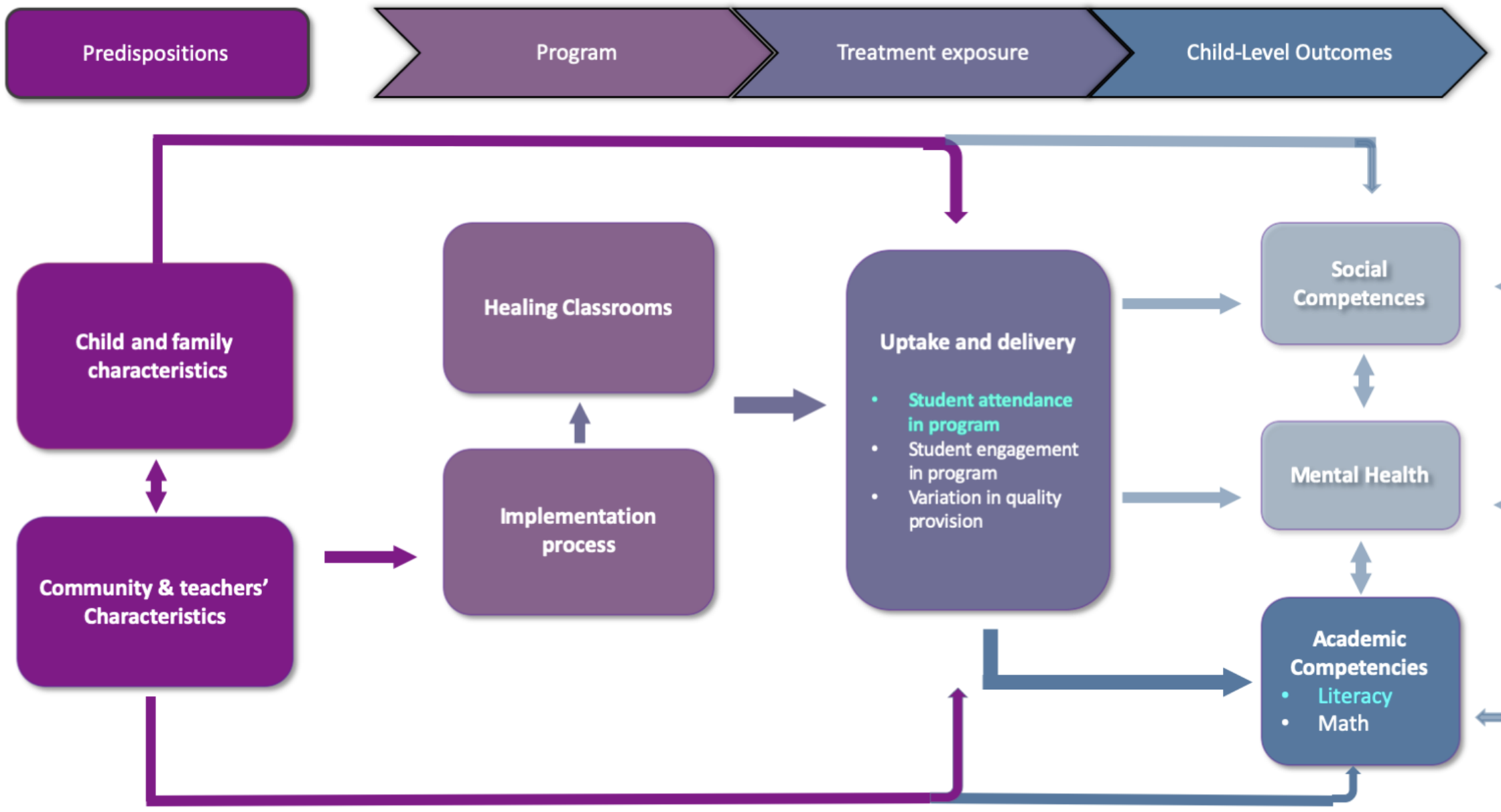

***Notes.*** *Depiction of child, family, and community characteristics on the left, as pre-treatment covariates. The program refers to elements of the intervention and its implementation, such as the types of mentoring and support that teachers received. Exposure heterogeneity is also considered. Mediating mechanisms are not part of the scope of this analysis and are excluded from the figure for simplicity. The child-level outcomes that the intervention targets are: (1) social competencies, (2) mental health, and (3) academic competencies, with the focus of this paper being children's literacy outcomes.*

## Sample

During the 2016-2017 school year, the 3EA partnership randomized refugee sites in Lebanon into 4 different treatment conditions: (1) **Two cycles (26 weeks) of basic HC remedial support,** (2) Two cycles (26 weeks) of HC remedial support with an added focus on targeted activities, (3) One cycle (10 weeks) of basic HC remedial support, and (4) One cycle (10 weeks) of HC remedial support with an added focus on targeted activities. Our sample of focus is *exclusively* the group of participants who were offered access to the *basic Healing Classrooms* ***remedial support*** over *2 cycles* of the program between 2016-2017. The sample of focus consists of 1,777 children (49.5% female) from 33 different refugee sites, aged 4 to 15 (M = 9.031, SD = 2.33) attending grades 1 to 9 (M = 2.722, SD = 1.715), and attendance rates averaging 35.4% (SD=24.29). Table 1 contains the descriptive statistics of basic demographics and baseline measures.

**Table 1** Basic descriptive statistics of analysis sample.

| variable | min | med. | mean | sd | max |
|---|---|---|---|---|---|
| Student’s sex at birth is female | 0 | 0.0 | 0.49 | 0.50 | 1 |
| Student’s age at baseline | 4 | 9.0 | 9.03 | 2.33 | 15 |
| School Grade | 1 | 2.0 | 2.72 | 1.71 | 9 |
| ASER Arabic score - baseline | 0 | 1.0 | 1.38 | 1.39 | 4 |
| Sum score of all letter recognition items - baseline | 0 | 16.2 | 21.81 | 20.78 | 50 |
| Sum score of all grapheme recognition items - baseline | 0 | 1.0 | 12.42 | 16.72 | 50 |
| Sum score of all oral reading items - baseline | 0 | 0.0 | 10.44 | 15.72 | 42 |
| Days of attendance to Healing Classroom | 0 | 37.0 | 35.43 | 24.29 | 80 |
| Deviation from expected age for grade | -5 | 1.00 | 1.31 | 1.52 | 7 |

***Notes.*** *Descriptives in the table are pooled after addressing missing data.*

**Measures**

One of our target outcome measures of children’s reading skills is a tool designed by the Annual Status of Education Report (ASER), which is intended to be collected nationwide and to measure children’s ability to read basic text in Arabic (ASER Centre, 2016; Vagh, 2009). The tool was administered by enumerators using a paper-based format. It consists of 4 types of exercises and rates children’s reading skills on a 5 point scale: 0- cannot recognize letters, 1- letter recognition, 2- word recognition, 3- short paragraph reading, and 4- longer paragraph reading. ASER is a useful tool in contexts with low resources that have a need for large scale reading screenings, due to its ease of implementation and the length of the assessment (Brown et al., 2021). The English version of the tool is presented in Figure A.1 of the online Appendix A, for reference.

Given high missingness rates and the bluntness of ASER, sub-scales from a different literacy tool, the Early Grade Reading Assessment (EGRA) (RTI-International, 2016), are also considered as outcomes of this study. Specifically, the sub-scales that more closely correspond to the reading skills targeted by the ASER reading tool were analyzed for comparison. These outcomes consist of the sum scores of 3 different sets of items that focus on: (1) letter recognition, (2) grapheme recognition, and (3) oral passage reading. The EGRA performance was collected through tablets by enumerators trained by the 3EA partnership. Students were given one minute to complete each type of task. The letter and grapheme recognition tasks were scored as the number of letters/graphemes correctly read (out of 60). The oral passage reading assessment consisted of the number of words from the passage that the student managed to read properly before 1 minute elapsed, and they were automatically scored 0 by the tablet if they missed all the words from the first sentence.

The unconditional changes from baseline to endline of the outcome measures are presented in Figure A.2 in online Appendix A . Note that the ASER and letter recognition outcomes exhibit a more pronounced increase (and distribution shift) from baseline to endline than the grapheme recognition and oral passage reading tools.

## Methods

### Missing Data

Several precautions were implemented to prevent missing data. Measurement tools were carefully selected and designed to reduce the participants' burden, mindful and culturally appropriate phrasing was used, data collection protocols allowed for several attempts to track absent children, and enumerators engaged in many individual efforts. Despite this, several factors made it virtually impossible to avoid missing data. These included: 1) the large number of assessments administered as part of the full 3EA study, 2) having to track down students on several separate waves for each assessment, and 3) challenging conditions to locate children (no formal identification, no stable residency, high mobility, security threats, etc.).

***Outcome non-response.*** The non-response rates for the ASER reading outcome measure in this analysis sample was 47.55% (n=845), while that for the 3 EGRA outcomes was 20.37% (n=362). This should make it clear to the reader that missing data played an important role during the research endeavors of the 3EA partnership amid understudied contexts that face protracted crises, and that the results presented from now on should be interpreted while considering these limitations. Non-response rates for each of the covariates selected for the impact models are presented in Table C.1 of online Appendix C.

***Addressing the missing data.*** The resulting missing data in the 3EA Lebanon 2016-2017 study was addressed comprehensively, using the Multiple Imputation framework (Rubin, 1987). We used a chained models (or “equations”) approach, commonly referred to as MICE (Van Buuren, Brand, Groothuis-Oudshoorn, & Rubin, 2006), within the R statistical programming language. This MICE approach creates a model for each variable with missing data, conditional on the other variables, and uses this to predict the missing values using all of the available information. We use Random Forests to model these relationships due to its ability to flexibly model nonlinear relationships, to accommodate a large number of predictors, and to identify each predictor's importance (Stekhoven & Bühlmann, 2012; Doove, Van Buuren, & Dusseldorp, 2014; Shah, Bartlett, Carpenter, Nicholas, & Hemingway, 2014). These methods were selected considering their ability to account for the uncertainty brought in by the missing data and to reflect it in the final estimates; with the goal to yield coherent, shareable, and comprehensive completed data sets to be used across different analyses, preventing assumptions about the non-response from being dependent on the model selected to address any one particular research question. The methods we used for convergence and imputation diagnostics of the outcome variables are described in sections 1 and 2 of online Appendix B.

***Outcome imputation diagnostics.*** The results of the imputation diagnostics (see online Appendix B) don't demonstrate meaningful disparities between the two groups (respondents and non-respondents), that might be left unexplained by the measured information that was used during the imputation process. While we can never rule out that data may be Not Missing At Random (NMAR), these diagnostics provide evidence that, given the information we have, our predictions for the missing data seem consistent with the data that were observed. The results from these diagnostics can help researchers feel more comfortable about the performance of the imputation models employed (Abayomi, Gelman, & Levy, 2008; Bondarenko & Raghunathan, 2016) and can offer peace of mind that there's no clear evidence against the assumption that the data are Missing At Random (MAR) (Rubin, 1987; Van Buuren, 2012). Although these results do provide us with a reasonable degree of confidence regarding the imputations for the EGRA scores, we feel less confident in the MAR assumption for the ASER non-response. The non-response rates in ASER were extremely high and also profoundly unbalanced between the treatment groups. It is likely that there was not enough representation among the limited pool of low-attendance respondents, of the non-respondents, to allow for successful bias adjustment (see *Limitations and Interpretation of Findings* and online Appendix B for details).

**Covariate Selection**

In the absence of a pristine randomization, covariates are typically required to satisfy the assumptions required for causal analysis. In theory, to identify a causal effect we simply need to properly adjust for all covariates that predict both treatment assignment and the outcome (confounders). We chose the covariates by considering 1) domain-knowledge (theoretical, research-based, and context-relevant variables incorporated in previous analyses by domain-experts (Kim et al. 2020; Tubbs et al., 2021)) and 2) a data-driven approach that minimizes researcher and cultural biases through the use of non-parametric models. In practice we always need to be mindful of the fact that it may not be possible to adjust for all such confounders.

Given the large number of survey items collected, the understudied nature of the context, and the migrant population of the 3EA study, a data-driven approach was instrumental to identify the items most important in predicting the outcome and treatment. This approach relies on a non-parametric recursive partitioning algorithm (Random Forests), which allows for flexible estimation of the models used to understand these relationships. A key feature of the algorithm is that it automatically incorporates non-linearities and interactions as long as they are sufficiently predictive of the outcome (Kursa, Jankowski, & Rudnicki, 2010; Kursa & Rudnicki, 2010; Ali et al., 2019; Degenhardt et al., 2019).

The final selection consists of 68 covariates that were deemed as relevant for at least one of the impact models. The final covariates include: baseline assessments of literacy and math skills, a measure of the children self-regulation in class, demographic variables of the teachers, the students, their households, and their migration history, variables related to the program training and implementation, as well as the teachers' experience, practices and perceptions. A complete explanation of the covariate selection process can be found in online Appendix C and the full list of covariates used in the model of each outcome in Table C.1.

**Analytic Approach**

***Assumptions for causal inference.*** To answer our research question causally we draw from the potential outcomes framework, proposed in Rubin (1978) and Rubin (2005). This framework allows us to think in terms of counterfactuals -- in this case what the outcome for each participant would have been under each of the different dosage assignments.

Say we allow $Y(0)$ to represent all of the post-treatment scores that would result for participants if they had been placed in a low attendance group. Then allow $Y(1)$ to represent what post-treatment scores would have been, had the same participants been placed instead in a high-attendance group. The average difference between these two sets of "potential outcomes" is defined as the average treatment effect (ATE). Unfortunately, we can only see one potential outcome for each person. We don't get to see Y(0) for those in the high attendance group or Y(1) for those in the low attendance group. This makes estimation challenging for causal inference. A pristine randomization of participants into the two groups would allow us to estimate this without bias. However, when the conditions of a pristine randomization are not present, then students might differ across exposure groups in ways that also predict their outcome. Addressing the selection bias this can create, requires stronger assumptions and more complicated methods.

The most critical assumption requires that we have measured all the pre-treatment covariates that predict both $attendance$ and the potential outcomes of the students. This assumption, often referred to as ignorability, can be formalized as,

$$Y(0), Y(1) \perp attendance | \mathrm{X}$$

If ignorability is satisfied, participants with comparable combinations of their covariates' values can be assumed to have comparable distributions in their potential outcomes and, thus, provide the necessary information to estimate reasonable counterfactuals for causal inference.

A second assumption, also known as common support (Hill & Su, 2013), essentially says that we have enough empirical counterfactuals between exposure groups (that share similar combinations of pretreatment covariates $X$), to make reasonable predictions about the potential outcomes $Y(0), Y(1)$ for all the students that are used to compute the average treatment effect. This assumption is expressed in terms of a non-zero probability of exposure for all subgroups of students that differ in their pretreatment characteristics:

$$0 < Pr(attendance|X) < 1$$

This assumption can be explored empirically to shed light on which observations may lack sufficient overlap. Inference for these observations should be interpreted with caution or avoided.

A final assumption, common to randomized experiments and observational studies, is the Stable Unit Treatment Value Assumption (SUTVA) and, for the purposes of this analysis, it means that the degree of attendance to the remedial classroom of any one child shouldn't affect the reading scores of other children. Even under a perfect randomization of attendance it is possible to imagine that children could collaborate outside the classroom in violation of this assumption. Making it important to acknowledge the possibility that the estimated treatment effects that can be obtained from a social study like this could be smaller than the true impact, as we cannot ensure that students with different attendance didn't help each other with their reading skills.

Beyond this, additional "parametric" assumptions are needed regarding our ability to accurately model the relationships between $attendance$, the pretreatment covariates $X$, and the potential outcomes $Y(0), Y(1)$. The plausibility of these assumptions depends on the modeling strategy used to estimate the treatment effect. Some approaches (such as linear models) rely more heavily on parametric assumptions than others. The major (and often implicit) assumption that all researchers make when estimating treatment effects is that the final model chosen captures the true relationships between the pretreatment covariates, the treatment variable, and $Y(0), Y(1)$.

***BART.*** The modeling approach used to estimate the impact of attendance in this study was Bayesian Additive Regression Trees (BART) (Chipman, George, & McCulloch, 2007, 2010). This method was used to model the relationship between the outcomes and the treatment and covariates. The resulting fit allowed us to to predict counterfactual outcomes for each participant from the analysis sample, which were then used to estimate the treatment effect.

BART is a semi-parametric modeling approach that combines elements of machine learning and Bayesian inference. It uses simple Regression Trees as a building block within a recursive partitioning algorithm that can naturally account for a large number of covariates, interactions, and other non-linearities in the relationship between the outcome variable, treatment, and covariates; that is, these features are only included in the model fit if they contribute sufficiently to variance explained. Embedding this algorithm within a Bayesian inferential framework allows for robust uncertainty quantification and helps prevent overfitting (one of the most popular concerns about tree-based models) by ensuring (through the prior specification) that each tree contributes only a small part to the overall fit (Hill, 2011; Hill et al., 2020).

In observational studies and in experiments where the conditions of pristine randomization can't be met, it is often crucial to appropriately adjust for all confounders to ensure ignorability. This can mean using strategies (matching, weighting, or flexible models) that can flexibly and simultaneously adjust for many covariates. BART is particularly well suited for this because it doesn't require strong assumptions about how the outcome relates to the covariates. Additionally, it is equipped to estimate counterfactuals for each participant in the analysis sample, along with the uncertainty of each counterfactual prediction. This feature allows us to better understand the observations in the sample for which we may lack *common* *causal support* (common support with respect to the true confounders, see Hill & Su, 2013) and where we should be more skeptical about our inference regarding the treatment effects.

The advantages of BART's performance when applied to causal inference settings, even in the absence of randomization properties, have been extensively demonstrated (Hill, 2011; Hahn, Murray, & Carvalho, 2017; Dorie, Hill, Shalit, Scott, & Cervone, 2019; Hahn, Dorie, & Murray, 2019) and tested under different configurations of Data Generating Processes (DGP) that have been calibrated to real world data. For more details on the model and its assumptions, please refer to (Hill et al., 2020). To account for the uncertainty induced by the missing data in the present study, for each outcome variable, 100 independent BART models were run, one on each imputed data set.

***Dichotomous treatment.*** First, for simplicity, the impact was assessed in terms of a more "traditional" dichotomous treatment variable: high attendance vs low attendance. This binary representation allows us to think in terms of comparisons between just two groups. The cutoff rule used to divide participants into the two groups was the median attendance value for the sample of focus. Raw differences in baseline measures and basic demographics between high and low attendance groups, after addressing non-response, are available in Table A.1 of online Appendix A for consultation. Children in the low attendance group have a higher chance of being male, are slightly older on average, and score slightly lower in most of the baseline scores (except for the oral passage reading) on average, compared to the children in the high attendance group.

The letter and grapheme recognition assessments show clear differences in their distributions across the low and high attendance groups (see Figure A.3 of the online Appendix A), while this is not the case for the ASER and oral passage reading scores; which are different from the former in that they both target the more advanced skills involved in full paragraph reading. After correcting for non-response there was a generalized adjustment of the outcome distributions towards lower scores, meaning that non-respondents were more likely to have lower scores than respondents; this was, however, not the case for ASER scores, and less so for the *oral passage reading* scores than for the other two tasks studied.

In addition to the covariates selected, the propensity score for belonging to the higher attendance group was included as a covariate in the model, as including it has been demonstrated to yield more accurate treatment effect estimates in many settings (Hahn et al., 2017; Hahn et al., 2019; Dorie et al., 2019; Hill et al., 2020).

The ATEs of moving from the low to the high attendance group were estimated for all imputed data sets using BART and then pooled using Rubin's rule to incorporate the uncertainty brought in by the missing data into the credible interval of the treatment effect estimate.

***Continuous treatment.*** Causal inference with a continuous treatment under the BART framework permits researchers to deal with non-experimental situations where the degree to which a participant is "treated" cannot be randomly assigned, and where confounders may play an important role. Within the framework of potential outcomes (unobserved counterfactuals) it allows us to obtain individual dosage trajectories for each participant, after conditioning on confounders, by estimating the outcome response at different dosage levels (Galagate, 2016), along with their uncertainty.

We use BART to estimate the average dosage-response function (ADRF) within a causal framework. This curve was produced by estimating counterfactuals for each student at different levels of attendance $a$=(0 days, 10 days, 20 days, etc), yielding an individual dosage-response function (DRF) for each participant. Next, the individual *response paths* were *pooled* at each dosage level $a$=(0 days, 10 days, 20 days, etc), across the DRFs of all students. This process was conducted separately for each of the 100 imputed data sets to visualize the uncertainty about the ADRF attributable to non-response.

***Common support.*** BART provides a strategy to assess the presence of common support for each observation (Hill & Su, 2013). It helps us understand whether we have enough information for each individual, about their counterfactual outcome, to justify estimating a causal effect for them. This approach assesses common support by comparing the uncertainty about the predicted outcomes for the counterfactual state (which we may or may not have information in the data about), relative to the uncertainty of the predictions of outcomes under the observed treatment conditions (which should have much more information about since this is based on observed data). This method tends to be less conservative for excluding observations than approaches that rely on comparisons of distributions of propensity scores, which can place too much importance on predictors of the treatment that may or may not be related to the outcome (Hill & Su, 2013). For a detailed description of the methods used to assess common support see Online Appendix F.

## Results

This section presents results both from our analyses of the impact of high versus low attendance on reading skills and from our analyses of how these impacts vary across days of attendance. We further disaggregate these effects by imputed datasets, and consider their variation across a range of potential moderators.

### High Versus Low Attendance Groups[1]

The results suggest a positive average effect of high versus low attendance on the most basic reading skills for the study participants, as measured by the letter (*ATE=2.14, SE=0.81*) and grapheme recognition (*ATE=2.92, SE=1.00*) sub-scales from the EGRA assessment. However, they also suggest there is no conclusive[2] impact, on average of moving from the lower attendance group to the high-attendance group on the ASER literacy scores (*ATE=0.00,*

[1] SEs throughout this manuscript are in fact the posterior standard deviation of each average treatment effect, which can be thought of as analogous to the standard errors under a frequentist framework, as they represent our uncertainty about the estimates.
[2] Meaning that zero is contained in the 95% Credible Interval of the ATE.

*SE=0.07*) and on the oral passage reading EGRA sub-scale (*ATE=0.60, SE=0.61*). Table 2 displays the ATE estimates, the 95% credible intervals, and two additional estimates related to the uncertainty brought by the missing data, known as the *fraction of missing information* about the estimate and the *relative increase in variance* due to non-response (Rubin, 1987; Van Buuren, 2012).

**Table 2** Average Treatment Effects: High vs Low attendance

| outcome | estimate | SE | 95% CI | fmi | riv |
|---|---|---|---|---|---|
| ASER reading | ***0.00*** | *0.067* | *[-0.13, 0.13]* | *0.304* | *0.432* |
| Grapheme recognition | ***2.14*** | *0.814* | *[0.55, 3.74]* | *0.109* | *0.120* |
| Letter recognition | ***2.92*** | *1.004* | *[0.95, 4.89]* | *0.135* | *0.154* |
| Oral passage reading | ***0.60*** | *0.607* | *[-0.59, 1.79]* | *0.251* | *0.331* |

***Notes.*** *CIs and posterior standard errors have been computed using Rubin's pooling rules, and already reflect the uncertainty brought in by the missing data. SE here is actually the posterior standard deviation of the ATE, which can be thought as analogous to the standard error under a frequentist framework, as it represents our uncertainty about the ATE estimate.*

The posterior distributions (Gelman et al., 2004) for the ATE for each of the reading outcomes in Figure D.1 of online Appendix D, reflect the impact of moving from the low to the high attendance groups, in each imputed data set. The plots show important variation in the impacts on *ASER* and *oral passage reading* measures across the different imputed data sets, illustrating the role that missing data plays in our ability to draw robust conclusions about these outcomes. The impacts on *letter* and *grapheme* recognition, on the other hand, show strong agreement across imputed data sets, which provides more certainty about the positive impacts of belonging to the high attendance group.

***Common support across high and low attendance groups.*** We found solid common support across the two attendance groups for all 100 imputed sets. In most of them no observation needed to be excluded from the analysis based on the common support rule proposed by Hill and Su (2013), and no more than 3 observations were excluded from any one imputed dataset.

***Impact heterogeneity.*** We assessed whether the effect of high versus low attendance varies by pretreatment scores, propensity scores, and the range of potential moderators that were available to us (at the individual, family, and community level). However, we found no compelling evidence of heterogeneous treatment effects across any of the factors considered. For details, please refer to Online Appendix E .

**Continuous Attendance and Incremental Impact**

All of the EGRA outcomes show a positive causal relationship between attendance in the HC and the outcomes. However, the dosage level at which the marginal impact seems to fade is around 40 days for the letter and grapheme recognition tasks, and around 20 days for the oral passage reading tasks. These results offer an additional explanation for why a clear impact wasn't found on the oral passage reading tasks, when comparing low and high attendance using a 37 day cut-off to split the groups.

Figure 2 displays the BART-estimated ADRFs for all four outcomes. The estimated ADRF for ASER is visibly flat, suggesting there is no meaningful causal relationship between attendance and the ASER reading scores. These results may be attributable to the issues discussed in the *interpretations of findings* section, however, and should be assessed carefully.

**Figure 2** Continuous attendance: BART ADRFs for different imputed data sets.

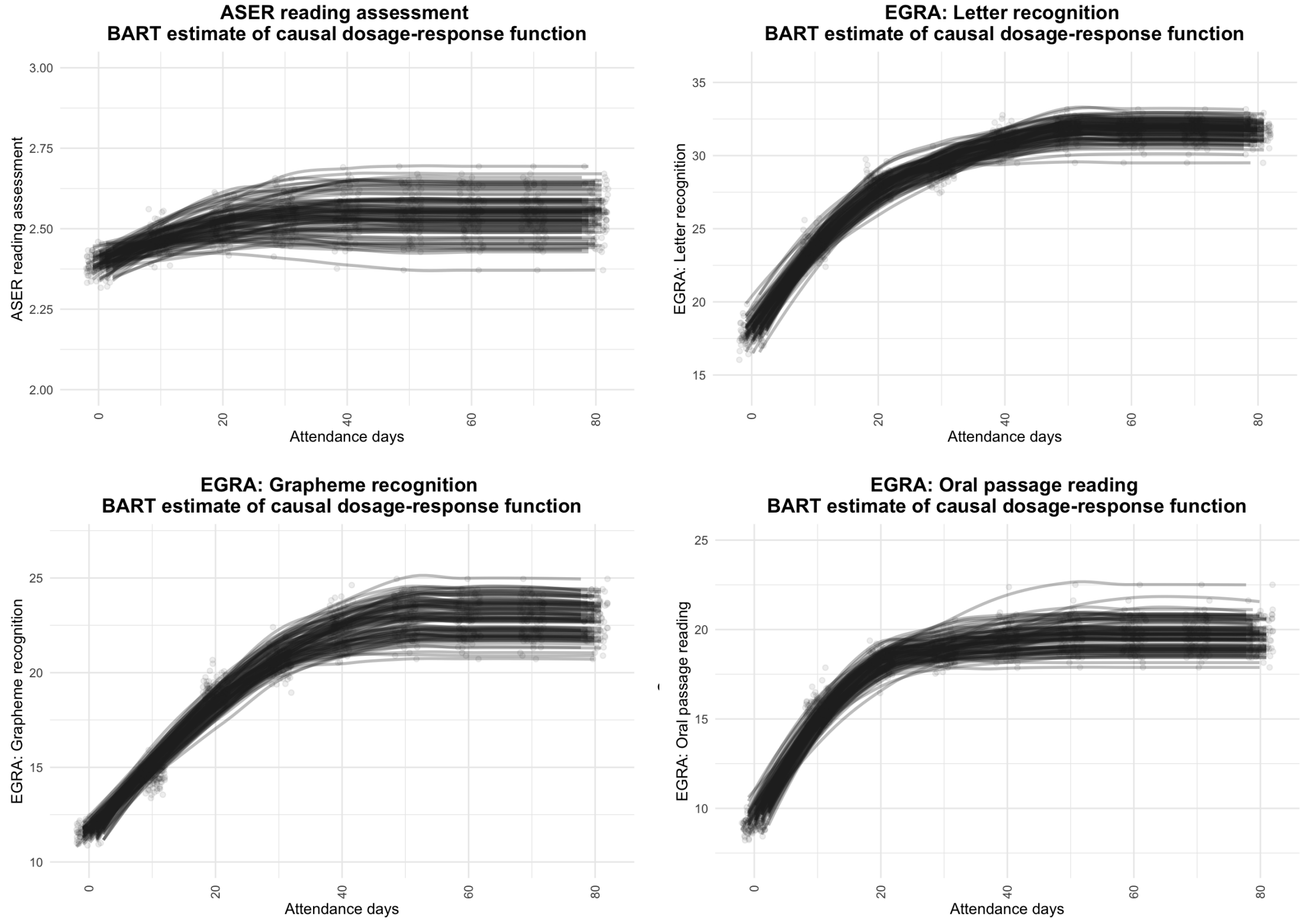

***Notes.*** *These figures illustrate the estimated causal relationship between attendance levels and each outcome, after controlling for all relevant covariates.*

***Common support at different levels of dosage.*** Under the relaxed rule our sample can be considered to have perfect common support, with all imputed sets agreeing that 100% of the analysis sample falls underneath the uncertainty threshold. See Figure F.1 in online Appendix F. Under the highly conservative cut-off, described in section 3 of online Appendix F, most of the sample retains sufficient causal support to reliably estimate counterfactuals and treatment effects across the attendance spectrum. Nonetheless, the ASER, grapheme recognition, and oral passage reading measures suggest that common causal support decreases for levels of attendance over 40 days, in many of the imputed data sets. On the other hand, when studying the impact on the letter recognition measure there seems to be less common causal support (around 70% of the sample having solid counterfactuals) at very low levels of attendance. See Figure F.2 in online Appendix F.

Depending on the goals and need for certainty and precision, the reader may choose to rely on either the more relaxed or conservative rule when interpreting the information that the ADRF provides about the impact of different levels of attendance to the remedial classroom on refugee children's reading scores.

***Attendance leaps.*** The estimated ADRFs suggest that the marginal impact of additional attendance days fades after reaching a certain threshold. To explore how much of the estimated effects of moving from 0 to 80 days of attendance is attributed to the first 37 days of exposure (median attendance observed), the posterior distributions of the average counterfactual predictions at 0, 37, and 80 attendance days were estimated and are displayed in Figure D.2 of online Appendix D.

All of the EGRA sub-scales show clear and large positive predicted effects of moving from 0 to 80 days of attendance. This is not the case for ASER, however, which has credible intervals that cross over zero. No clear impact was found for any of the outcomes of moving from 37 to 80 days of attendance within the **26 weeks** that the program lasted, which suggests that the vast majority of the gains from this remedial program on the reading scores of Syrian refugee children can be attributed to the first 37 days of registered attendance. The reader should be mindful, however, that these results do not necessarily extend to what the effect would be of *sustained attendance* over a longer period of time.

Heterogeneity across participants with different *observed* attendance (see Figure G.1, Figure G.2, and Figure G.3 of online Appendix G) was explored for the estimated treatment effects of moving: (1) from 0 to maximum attendance, (2) from 0 to median attendance, and (3) from median to maximum attendance, separately. No heterogeneity of impact was found across children with different factual levels of attendance. This means that children who were low or high attending don't appear to be different, in terms of the benefits they can harvest from the program, had they been able to attend at the same rates. Note that this is only true for the reading skills studied here, and may not be the case for the benefits on other types of outcomes that were targeted by the intervention.

**Non-Response Cautionary Note**

Missing data played an important role in this study and the sample of respondents appears to be systematically biased prior to adjusting for non-response. Particularly, because non-response in the outcome variables was closely tied to attendance (the treatment variable of interest in this report), the treatment effect estimations that can be obtained from such a biased sample are expected to differ from the estimates after addressing the missing data.

When looking only at the sample of respondents for each outcome, the ATE on all of the outcome measures comes up as centered around zero. The impact estimates for the likely biased sub-samples of respondents for each outcome were computed for transparency and can be consulted in Table B.3 of online Appendix B. As a reminder, the sample loss when estimating the ATEs without using the outcome imputed data would be close to 50% for ASER and around 20% for the EGRA sub-scales. The contrast of these results suggests that the children that were less likely to be located during the post-treatment assessments, could have been those that had the most to gain from the remedial support.

## Limitations and Interpretation of Findings

A combination of factors conspired to create differences in data quality across the outcomes reviewed here. These differences play a role in the ability of outcome measures to capture improvements in children’s reading skills, especially in the ASER measure.

**Differences in Implementation**

Differences in the implementation of each assessment tool and in the data collection protocols across outcomes, yielded heterogeneity in data quality; which is also reflected in the differences in impact estimates and their uncertainty. Additionally, some of the assessment protocols of the tools used in this analysis could have contributed to exacerbate zero-inflation and artificially obscure treatment effects.

First, there were substantial differences in how ASER was approached during data collection. ASER was captured in *administrative* data and was not explicitly designed as a measure for research purposes. Rather, ASER was used by the IRC for practical and quick-decision-making during study implementation. In particular, it was used as a screening tool to place students into different remedial levels and was collected as part of a separate *data collection mechanism* implemented independently by the IRC. As such, this data collection did not adhere to the partnership protocols used for the collection of the central research data. One of the most significant differences in the data collection protocol for ASER was a policy of assessing children on “one day and one day only,” with no follow-up set in place for the children that were missing. In contrast, for EGRA, additional efforts were implemented to track down the missing children and to maximize representation of those that were less likely to be present on a given day. This important difference in data collection is directly intertwined with the concerns that are highlighted in the *hidden ASER bias section of* online Appendix B. Additionally, differences in the data collection protocols could have made the ASER scores more prone to data entry issues.

### Differences in Measure Quality

***Discrete nature of ASER.*** Contrary to the EGRA sub-scales, which allow tracking improvements within each type of reading tasks, the ASER assessment is *discretized* and provides only information of full (assessment-based) proficiency on each type of task by grading students on a 5-point scale from 0 to 4. If attendance to the program over the course of the 26 weeks was more likely to impact the most basic reading tasks than it was to impact the higher level reading tasks (which is what the analysis of the different EGRA sub-scales suggest), then the ASER instrument wouldn’t be able to capture and reflect such impact.

***Zero-inflation.*** EGRA sub-scales are not exempt from data quality concerns. Stop rules were implemented for these outcomes before computing their scores to minimize assessment burden. If students didn't give a correct answer within the first 10 items, then they weren't given an opportunity to respond to the remaining items of that section, and their responses would get automatically populated as incorrect. This practice can artificially zero-inflate the collected outcomes and contribute to obscuring the treatment effects. The repercussions on the *oral passage reading* outcome from this issue are twofold. First, the assessment skip-rules made it so that the opportunity to be presented with the task depended on the score that the child received on both the *letter recognition* and *grapheme recognition* tasks. Second, even when the task was presented to them, they would be prevented from responding to all of the following *oral passage reading* items if they were to miss the very first words from the passage.

**Missing Data**

Tracking and identification of participants presented many challenges in a conflict-affected setting where the population constantly relocates and often lacks any form of official identification. Relentless efforts were conducted to retrieve the data of children with incorrect or missing IDs through data verification, to minimize the loss of collected data, and principled methods were used to address missing data after data verification was completed. However, because of the real-world challenges to track down children throughout several waves of data collection in such a volatile context, missing data from this study is still not negligible. Despite encouraging results obtained from diagnostics checks, we cannot rule out the possibility of a hidden NMAR bias due to unmeasured information.

Although we find no direct evidence of obvious NMAR, there is a chance that they bias our estimates of the impact of Healing Classrooms on the ASER outcome measure (see Online Appendix B). Nonetheless, we believe that this impact evaluation provides crucial information about the intervention and about the challenges of conducting rigorous research under very precarious and volatile field conditions.

Furthermore, Syrian refugee children could be affected by unmeasured factors that could impact their ability to be present for all measured assessments ***as well as*** their attendance to the program. Further efforts to track down non-respondents and identify unmeasured reasons for missing data could help inform future data collection processes in similar contexts and enhance the chances to prevent NMAR biases. It remains for future research efforts to develop better identification and contact systems during data collection that are appropriate for the differences in culture and in the physical reality that families and surveyors face in conflict-affected regions.

### Cultural Bias

Over and above these constraints, this study entails limitations that are inherent in conducting research in an understudied context and when approaching it from frameworks that have been developed under the conditions and perspective of the westernized regions of the world. As scholars external to the region and population of study, we acknowledge the risk that our framework and methods could still be insufficient for the needs of a study under a different cultural and practical context.

## Conclusions and Discussion

This paper makes several important contributions to the field. We discuss these contributions here, summarize our major findings, and discuss the implications.

### Contributions

Educational interventions remain understudied in conflict-affected and migrating populations. In fact, most of the evidence that causally links attendance and academic outcomes comes from the United States, but little is known about these impacts outside that narrow scope. This paper contributes to the literature by studying an educational intervention serving some of the most at-risk children in the world -- Syrian refugee children in Lebanon.

Furthermore, this paper broadens our understanding of attendance by focusing on *out-of-school* SEL-infused remedial support programs, and shows how increased exposure to a safe and supportive learning environment can benefit the academic development of children in crisis contexts.

Our third contribution is methodological. We used more advanced statistical methods (BART for causal inference and multiple imputation relying on random forests) than are typical in most studies of attendance as a way to minimize the strong assumptions that are often necessary when making causal claims, particularly when missing data are involved. We also were careful to review whether there was sufficient information in data to support our causal claims. While we always need to be careful when estimating causal effects, we invested extraordinary efforts to use all of the information available (by not discarding observations with missing data) and to relax the assumptions required for causal inference (e.g. parametric and overlap assumptions) to the extent that was possible.

In short, we are aware of no prior study of the impact of attendance in *out-of-school* SEL-infused remedial support programs for refugee children that has so rigorously accounted for selection bias and the many data challenges that are prevalent in crisis contexts.

**Impact Conclusions**

***High versus low attendance.*** High attendance led to improvements relative to low attendance for the most basic reading scores from the EGRA assessment: *grapheme recognition* and *letter recognition*. No conclusive impact of attendance was found on the more advanced task of *oral passage reading* from moving from the low to the high attendance group.

***Continuous dosage.*** The average response of reading skills to different levels of attendance, shows a positive and growing impact on the measured reading skills, except for the ASER assessment which suffers important limitations due to non-response. This positive impact of attending more days *within the 26 weeks offered*, displays diminishing returns and becomes stagnant after crossing 40 days of attendance for *letter* and *grapheme recognition*; while for *oral passage reading* the marginal returns of attending start disappearing after 20 days of attendance during the 26 week time frame. The considered measures for common support suggest that even under highly cautious rules, most of the sample retains sufficient causal support to estimate adequate counterfactuals and treatment effects across the possible attendance spectrum.

***Attendance leaps.*** Our exploration of counterfactuals at large leaps of attendance suggests that the effect of moving from zero attendance days to 37 days, during the 26 weeks studied, is substantial for all measures (except for ASER) and it represents a potential increase, on average, of over 10 points in reading outcome scores.

Findings should be interpreted carefully and always under consideration of the limitations presented by the non-response. Moreover, each refugee population is culturally different, quickly changing and adapting over time, and their interaction with each host community can likely impact how children develop and learn.

## Methodological Conclusions

An important takeaway from our interpretation about these findings is that research decisions can have substantial repercussions on what can be learned about an intervention, beyond the analysis method chosen. Such decisions include: 1) the measurement tool used, 2) the tool administration protocols, 3) the attempts to track down non-respondents during each wave, 4) the instrumental information in the assessments that allows for data verification, and 5) choices regarding how to address missing data (as opposed to efforts to only prevent missing data).

**Summary**

We conclude that *attending* SEL-based remedial support shows promise in improving basic reading outcomes in Syrian refugee children, even after only 26 weeks of offered access. We encourage researchers to consider how to improve future studies especially with regard to (a) identifying unmeasured causes for non-response to prevent the risks of NMAR biases in contexts where missing data will continue to be a certainty due to migration, and (b) thoroughly supporting the work required for mindful selection and implementation of appropriate assessment tools, for dedicated tracking of participants and data verification, as well as for adequately addressing non-response.

It is crucial to grow rigorous research efforts on refugee children's development across contexts with an understanding that blunt generalizations across refugee populations can be inadequate. It is also necessary to work towards establishing mechanisms for research to reach (and be used by): the refugee populations themselves, the international agencies that are implementing these efforts, and the policy makers from the host communities.'